\documentclass[12pt,prd,aps,amssymb,amsmath,tightenlines,showpacs]{article}
\usepackage[utf8]{inputenc}
\usepackage[a4paper,
top=3cm,bottom=3cm,left=1.5cm,right=1.5cm]{geometry}
\usepackage{amssymb}
\usepackage{amsmath}
\usepackage{amsthm}
\usepackage{tipa}
\usepackage{latexsym} 
\usepackage{graphicx}
\usepackage{slashed}
\usepackage{bbold}
\usepackage{cancel}
\usepackage{comment}
\usepackage{physics}
\usepackage{color}
\usepackage{graphicx,epsfig,color}
\usepackage{comment}
\usepackage{cite}
\usepackage{tikz}
\usepackage[compat=1.1.0]{tikz-feynman}
\usetikzlibrary{decorations.markings}
\usepackage{color}
\usepackage{graphicx,epsfig,color}
\usepackage[dvipsnames]{xcolor}
\usepackage{hyperref}
\hypersetup{colorlinks, linkcolor=Magenta,citecolor=Cerulean}

\tikzfeynmanset{
	photon to bigotimes/.style={
		/tikz/postaction={
			decorate,
			decoration={
				markings,
				mark=at position 0.84 with {
					\node[anchor=west] {$\bigotimes$};
				}
			}
		}
	}
}

\tikzfeynmanset{
	photon to bigotimes2/.style={
		/tikz/postaction={
			decorate,
			decoration={
				markings,
				mark=at position 0.80 with {
					\node[anchor=west] {$\bigotimes$};
				}
			}
		}
	}
}

\newcommand{\be}{\begin{equation}}
\newcommand{\ee}{\end{equation}}
\newcommand{\bea}{\begin{eqnarray}}
\newcommand{\eea}{\end{eqnarray}}
\newcommand{\vv}{``}

\numberwithin{equation}{section}

\begin{document}
	\graphicspath{{FIGURE/}}
	\topmargin=-2cm

\begin{flushright} {\small KA-TP-13-2026}  \\ \end{flushright}
\begin{center} 

{\Large{\bf Quantum gravity and spectral running cutoff}}

\vspace*{0.8 cm}

	{\small 
C. Branchina$^{\S,\,a},$ \let\thefootnote\relax\footnote{$^\S$carlo.branchina@kit.edu}	
V. Branchina$^{\dagger,\,b,c},$ \let\thefootnote\relax\footnote{$^\dagger$branchina@ct.infn.it}F. Contino$^{\ddagger,\,d,e},$ \footnote{$^\ddagger$f.contino@ssmeridionale.it}\,R. Gandolfo$^{\star,\,b,c},$ \footnote{$^\star$riccardo.gandolfo@ct.infn.it }A. Pernace$^{*,\,b,c}$ \footnote{$^*$arcangelo.pernace@ct.infn.it}}
		
				\vspace*{0.4cm}
		
	{\it	
		
		{\footnotesize{${}^{a}$\it Institute for Theoretical Physics, Karlsruhe Institute of Technology, \protect \\Wolfgang-Gaede-Str. 1,76131 Karlsruhe, Germany}}\\
		
		\vskip 5pt
		
		${}^b${\footnotesize Department of Physics, University of Catania, Via Santa Sofia 64, I-95123 Catania, Italy}
		
		\vskip 5pt
		
			${}^c${\footnotesize INFN-Catania,
						Via Santa Sofia 64, I-95123 
						Catania, Italy}
				
				\vskip 5pt
				
				${}^d${\footnotesize Scuola Superiore Meridionale, Largo San Marcellino 10, 80138 Napoli, Italy}
				
				\vskip 5pt
				
				${}^e${\footnotesize INFN-Napoli, Complesso Universitario di Monte S. Angelo, Via Cinthia Edificio 6, 80126 Napoli, Italy}}

\vskip 1cm

\setcounter{footnote}{0}

{\bf\small Abstract}
\end{center}

{\small
\noindent
We have recently shown that a natural way to implement the Wilsonian paradigm in  gauge theories is through the introduction of a \vv spectral cutoff'', a cut on the eigenvalues of the covariant Laplacian, pointing out that this provides the route toward the renormalization group (RG)  construction. Here we apply this idea to quantum gravity, resorting to two realizations of the spectral running cutoff: \vv hard'' and \vv smooth''. We derive the RG equations for the Newton and cosmological constant and find the RG pattern of the asymptotic safety scenario, with a non-Gaussian  UV-attractive fixed point. 
}

\section{Introduction}

Wilson's lesson\cite{Wilson:1971bg,Wilson:1971dc,Wilson:1971dh,Wilson:1973jj, Wegner:1972ih,Wilson:1974mb} provides one of the deepest insights in quantum field theory.  The general idea is that the  couplings that appear in the action $S_k$ at a given scale $k$ encode the effect of degrees of freedom that have already been integrated out. The renormalization group (RG) flow is obtained by lowering the scale and progressively integrating quantum fluctuations in infinitesimal shells. 

While in non-gauge theories and in theories formulated on flat spaces the idea of infinitesimal shells is naturally implemented in terms of  momentum shells, in gauge theories, in gravity (where the  notion of momentum cannot even be globally introduced) and in general in theories formulated on non-flat spaces, the issue is more delicate. The introduction of a hard cutoff on momenta is not a covariant operation and can easily generate violations of Ward, Slavnov-Taylor or diffeomorphism identities. 

We have recently shown  that the natural way to implement the Wilsonian paradigm in  gauge theories is through the introduction of a \vv spectral cutoff'', a cut on the eigenvalues of the covariant Laplacian\,\cite{Branchina:2026vvk} (similar ideas were already expressed in\,\cite{Branchina:2003ek}). The Wilsonian idea is then realized in a form that appears physically appropriate for a theory with gauge redundancy and/or diffeomorphism invariance. Moreover, we pointed out in \cite{Branchina:2026vvk} that this is how \vv infinitesimal shells'', that are at the basis of the RG construction, should be introduced in this context\footnote{These ideas have been recently applied to the simple  example of a scalar theory defined on $S^3$ \cite{Bonanno:2026bbb}.}. 

In the present work we apply to quantum gravity the idea we put forward in \cite{Branchina:2026vvk}. Resorting to the Einstein-Hilbert truncation for the gravitational action, we take a spherical background and define the shells in terms of the eigenvalues of the corresponding covariant Laplacian operator.
We then consider two different realizations of the running spectral cutoff $k$ (smooth and hard), and derive the RG equations for the running Newton and cosmological constant. Anticipating on the results, we find that the RG flow presents the pattern of the asymptotic safety (AS) scenario. A previous study of ours led to a different outcome \cite{Branchina:2024lai}. The critical element at the root of the diversity in the results is the identification of the running scale in terms of a spectral cut\cite{Branchina:2026vvk}, which in turn determines the physically meaningful separation of scales encoded in the theory.

The possibility of an AS scenario for quantum gravity was originally hinted by Weinberg \cite{Weinberg1979}. A concrete implementation was first given by Reuter \cite{Reuter:1996cp} within the framework of the functional RG formalism,  and later developed in \cite{Granda:1997xk,Granda:1998wn,Falkenberg:1996bq,Reuter:2001ag,Lauscher:2001ya,Litim:2003vp,Reuter:2012id,Niedermaier:2006wt,Bonanno:2004sy}. For instance, in the formulation implemented through the effective average action $\Gamma_k$ \cite{Wetterich1993} (with $k$ the running scale) one adds to the inverse propagator $\Gamma_k^{(2)}$ a regulator $R_k$, and the flow is governed by a trace involving $(\Gamma_k^{(2)}+R_k)^{-1}k\partial_k R_k$. Our approach differs from these realizations.  
We believe that it is more directly Wilsonian in that it is tied to the \vv Wilsonian effective action'' $S_k$ rather than the \vv average effective action'' $\Gamma_k$ for which some conceptual issues are yet to be sorted out, notably the reconstruction problem \cite{Basile:2024oms}. Furthermore, in our approach the running scale $k$ is introduced through the invariant eigenvalues of the Laplacian operator, which directly provide the intrinsic hierarchy of scales.

In this respect, it is worth to recall here that in the realization of the
AS scenario within the functional RG program some important questions are still open. It has not yet been established whether the dimension of the UV critical surface is stable against the addition of higher order operators, though some truncations beyond Einstein-Hilbert have already been considered \cite{Lauscher:2002sq,Codello:2007bd,Machado:2007ea,Codello:2008vh,Benedetti:2009rx,Falls:2014tra,Hamada:2017rvn,Falls:2020qhj}. Another relevant question concerns the dependence of the critical exponents on the regulator. Finally, while in early literature it was advanced that
the product $\lambda^{\text{\tiny AS}}_* g^{\text{\tiny AS}}_*$ does not depend on the regulator \cite{Lauscher:2001ya,Bonanno:2004sy} ($(\lambda^{\text{\tiny AS}}_*, g^{\text{\tiny AS}}_*)$ is the AS fixed point, with $\lambda$ and $g$ the dimensionless cosmological and Newton constant, respectively), in more recent literature this does not appear to be the case \cite{Litim:2003vp,Basile:2024oms}. We hope that our work might be of help in clarifying some of these or other open questions.

The rest of the paper is organized as follows. In section\,\,\ref{sec:oneloop} we prepare the tools for the RG analysis. In section\,\,\ref{sec: proper-time} we derive the RG equations for the Newton and cosmological constant ($G_k$ and $\Lambda_k$, respectively) using a smooth spectral cutoff, and study the fixed points and critical exponents. In section\,\,\ref{sec: Hard cutoff} we derive the RG equations for $G_k$ and $\Lambda_k$ using a hard spectral cutoff. Then, passing to dimensionless couplings, we derive the beta functions and analyze the fixed points and critical exponents. Section\,\,\ref{sec:conclusions} is for the conclusions and outlooks.

\section{General set-up}
\label{sec:oneloop}

In the present section we briefly recall  some steps of the one-loop calculation of the effective action that are relevant to the following analysis. Let us consider the Einstein-Hilbert truncation
\begin{equation}
	S[g]= {1\over 16\pi G}\int \dd^4x\sqrt{g}\,\left(-R+2\Lambda_{\rm cc}\right),
	\label{eq:EHaction}
\end{equation}
and take as background the metric $g_{\mu \nu}^{(a)}$ of a sphere of radius $a$. As in this case $\int \dd^4x\sqrt{g^{(a)}}={8\pi^2\over 3}a^4$ and $R(g^{(a)})={12\over a^2}$, the action \eqref{eq:EHaction} takes the form
\begin{equation}
	S^{(a)}\equiv S[g^{(a)}]={\pi\Lambda_{\rm cc}\over 3G}a^4-{2\pi\over G}a^2.
	\label{eq:classicalSphere}
\end{equation}

Following \cite{TaylorVeneziano1990}, where the off-shell one-loop effective action is obtained resorting to the Vilkovisky-DeWitt geometrical approach\cite{Vilkovisky:1984st,DeWitt:2003pm}, we write the one-loop contribution to the effective action as 
\begin{align}
	\delta S^{\rm 1l}
	= -{1\over 2}\log
	{\det_1\,\big[(-\Box^{(1)}-3/a^2)/\Lambda^2\big]\,
		\det_2\,\big[(-\Box^{(0)}-6/a^2)/\Lambda^2\big]
		\over
		\det_0\,\big[(-\Box^{(2)}-2\Lambda_{\rm cc}+8/a^2)/\Lambda^2\big]\,
		\det_2\,\big[(-\Box^{(0)}-2\Lambda_{\rm cc})/\Lambda^2\big]}\,,
	\label{eq:oneloopdet}
\end{align}
where $-\Box^{(s)}$ are the spin-$s$ Laplace-Beltrami operators, and the index $i$ in $\det_i$ indicates that the product over eigenvalues starts from $\lambda^{(s)}_{s+i}$. The corrections to $\Lambda_{\rm cc}/G$ and $1/G$ are extracted from the coefficients of $a^4$ and $a^2$, respectively\footnote{The scale $\Lambda$ in\,\eqref{eq:oneloopdet}, that takes care of the dimension of the fluctuation operators, is the Wilsonian UV cutoff of the theory. It arises from the phase space path integral after integration over the conjugate momenta\,\cite{Branchina:2026vvk}. In the case of gravitational theories, this integration also gives rise to the non-trivial local factors of the Fradkin-Vilkovisky measure \cite{Fradkin:1973wke}. These are non-covariant factors sometimes claimed to break diffeomorphism invariance. As shown in \cite{Fradkin:1973wke,Fradkin:1975sj}, and more recently in \cite{Branchina:2025lqw}, it is actually the opposite:  diffeomorphism invariance is guaranteed thanks to the presence of these terms. The reason why we do not write them explicitly in\,\eqref{eq:oneloopdet} is that, as shown in \cite{Fradkin:1976xa}, they are cancelled by equal (and opposite in sign) terms coming from the logarithm of the fluctuation determinants.} (see \eqref{eq:classicalSphere}).

The eigenvalues $\lambda_n^{(s)}$ of the Laplace-Beltrami operators on a four-sphere of radius $a$ and the corresponding degeneracies $D_n^{(s)}$ are
\begin{equation}
	\lambda^{(s)}_n={n^2+3n-s\over a^2},
	\qquad
	D^{(s)}_n={2s+1\over 3}\left(n+{3\over2}\right)^3
	-{(2s+1)^3\over 12}\left(n+{3\over2}\right),
	\label{eq:sphereSpectra}
\end{equation}
with $n=s,s+1,...$ .

Following\,\cite{Branchina:2026vvk}, we now introduce the spectral running cut that defines the \vv infinitesimal RG step''. To this end we observe that 
 in \eqref{eq:oneloopdet} each of the fluctuation  operators   has the form
\begin{equation}
\mathcal{O}_{s,\alpha}= -\Box^{(s)}+\alpha\,, 
\label{eq:genericLaplaceOperator}
\end{equation} 
and that the eigenvalues $\lambda_n^{(s)}$ of the Laplace-Beltrami operators $-\Box^{(s)}$ with generic spin $s$ are related to the eigenvalues  $\lambda_n^{(0)}$ of $-\Box^{(0)}$ by the simple relation
\begin{equation}
\lambda_n^{(s)}=\lambda_n^{(0)}-{s\over a^2} .
\label{eq:spinScalarShift}
\end{equation}
As stressed in\cite{Branchina:2026vvk} and in the Introduction, the natural way to implement the Wilsonian paradigm in  gravity (more generally in gauge theories) is through the introduction of a \vv spectral cutoff'' $k$ on the eigenvalues of the covariant Laplacians. This is an invariant way to introduce a separation between IR and UV. In gauge/gravitational theories in fact this separation cannot be obtained with a momentum cutoff since the latter does not respect gauge/diffeomorphism invariance\footnote{The  notion of momentum cannot even be globally introduced in the case of a curved space.}. Therefore, thanks to\,\eqref{eq:spinScalarShift}, we define the RG step resorting to\,\eqref{eq:oneloopdet} and calculating the determinants of the restriction of the fluctuation operators to the \vv infinitesimal shell'' given by 
\begin{equation}
(k-\delta k)^2\lesssim \lambda_n^{(0)}\lesssim k^2\,.
\label{eq:shell}
\end{equation}
In other words, we are not using the spin-dependent shift to define different cutoffs in different spin sectors: the Wilsonian running cutoff $k$ is identified with the scalar Laplace-Beltrami spectral scale.

In the coming sections we implement the RG strategy  outlined above in two ways.  In section \ref{sec: proper-time} we introduce a smooth spectral floating cutoff, while in section \ref{sec: Hard cutoff} we consider its hard counterpart.

\section{Smooth spectral cutoff}\label{sec: proper-time}

We now turn to the realization of the RG strategy presented in the previous section introducing a smooth running spectral scale. To this end, we resort to the proper-time representation, and recall that, for a generic operator $\mathcal O$ with eigenvalues $\lambda_n$ and degeneracies $D_n$, $\log\det \mathcal O$ is given by  
\begin{equation}
	\log \det \mathcal O\big|_\Lambda
	=-
	\sum_n D_n
	\int_{1/\Lambda^2}^{\infty}{\dd \tau\over \tau}
	\exp\left[-\tau\lambda_n\right] .
	\label{eq:ptTraceGeneric}
\end{equation}
In this framework, the RG strategy  is realized replacing each of the $\log\det\mathcal{O}_{s,\alpha}$ in\,\eqref{eq:oneloopdet} with
\begin{equation}
\Big(\log\det\mathcal{O}_{s,\alpha}\big|_{k-\delta k,k}\Big)_{\rm smooth} =-\sum_n D_n^{(s)}
\int_{(k^2-s/a^2+\alpha)^{-1}}^{((k-\delta k)^2-s/a^2+\alpha)^{-1}}{\dd \tau\over \tau}
\exp\left[-\tau\left(\lambda_n^{(0)}-{s\over a^2}+\alpha\right)\right] .
\label{eq:ptTrace Os}
\end{equation}
In fact, the lower bound in the proper time integral implements the suppression of the eigenvalues
\begin{equation}
\lambda_n^{(s)} = \lambda_n^{(0)}-{s\over a^2}+\alpha\gtrsim k^2-{s\over a^2}+\alpha,
\end{equation}
while the upper bound ensures the suppression of the eigenvalues 
\begin{equation}
	\lambda_n^{(s)} = \lambda_n^{(0)}-{s\over a^2}+\alpha\lesssim (k-\delta k)^2-{s\over a^2}+\alpha.
\end{equation}

Replacing each of the determinants in\,\eqref{eq:oneloopdet} according to\,\eqref{eq:ptTrace Os}, expanding to first order in $\delta k$, and taking the limit $\delta k\to 0$, we obtain the flow equation for the running action $S_k$
\begin{align}
       k{\partial S_k\over\partial k}
        =\frac12 k
        \,{\partial\over\partial k} \sum_{n=2}^{\infty}
        \Big[
        & D_n^{(0)}
        \int_{(k^2-2\Lambda_k)^{-1}}^{\infty}{\dd \tau\over \tau}
        \exp\left(-\tau\left(\lambda_n^{(0)}-2\Lambda_k\right)\right)
         \nonumber \\
        + & D_n^{(2)}
        \int_{(k^2+6/a^2-2\Lambda_k)^{-1}}^{\infty}{\dd \tau\over \tau}
        \exp\left(-\tau\left(\lambda_n^{(0)}+6/a^2-2\Lambda_k\right)\right) \nonumber \\
        - & D_n^{(1)}
        \int_{(k^2-4/a^2)^{-1}}^{\infty}{\dd \tau\over \tau}
        \exp\left(-\tau\left(\lambda_n^{(0)}-4/a^2\right)\right) \nonumber \\
        - & D_n^{(0)}
        \int_{(k^2-6/a^2)^{-1}}^{\infty}{\dd \tau\over \tau}
        \exp\left(-\tau\left(\lambda_n^{(0)}-6/a^2\right)\right)\Big]\,,
        \label{eq:ptRG}
\end{align}
where in the right-hand side the derivatives with respect to $k$ are taken keeping the couplings $G_k$ and $\Lambda_k$ fixed.
Inserting in the left-hand side of \eqref{eq:ptRG} the Einstein-Hilbert truncation
\begin{equation}
        S_k^{(a)}={\pi\Lambda_k\over 3G_k}a^4-{2\pi\over G_k}a^2\,,
        \label{eq:EHtruncationRunningPT}
\end{equation}
expanding the right-hand side for large $a$, and matching the coefficients of $a^2$ and $a^4$, we get the flow equations for $1/G_k$ and $\Lambda_k/G_k$
\begin{align}
        k {\dd\over\dd k}\left({1\over G_k}\right)
        &={k^2\over 6\pi}
        \left[
        1+
        {9k^2+12\Lambda_k\over k^2-2\Lambda_k}
        \exp\left({2\Lambda_k\over k^2-2\Lambda_k}\right)
        \right]  
        \label{eq:ptFlow1overG}
        \\
        k {\dd\over\dd k}\left({\Lambda_k\over G_k}\right)
        &={k^2\over \pi}
        \left[
        -2k^2+3\left(k^2-2\Lambda_k\right)
        \exp\left({2\Lambda_k\over k^2-2\Lambda_k}\right)
        \right] .
        \label{eq:ptFlowLambdaOverG}
\end{align}

The presence in \eqref{eq:ptFlow1overG} and \eqref{eq:ptFlowLambdaOverG} of   terms that depend on $k^2 - 2\Lambda_k$  indicates that for positive values of the cosmological constant the RG flow may hit a singularity at a finite value of the running scale $k$. This is related to the well-known IR instability associated with the propagator in de Sitter space\cite{Allen:1986ta,Polyakov:1982ug,Polyakov:2000fk,Jackiw:2005yc}.

The system \eqref{eq:ptFlow1overG}-\eqref{eq:ptFlowLambdaOverG} can be translated in the RG equations for $G_k$ and $\Lambda_k$,
{\small 
\begin{align}
        k{\dd G_k\over\dd k}
        &=- {k^2 G_k^2\over 6\pi}
        \left[
        1+
        {9k^2+12\Lambda_k\over k^2-2\Lambda_k}
        \exp\left({2\Lambda_k\over k^2-2\Lambda_k}\right)
        \right] \label{eq:ptFlowGdimensionful} \\
        k{\dd \Lambda_k\over\dd k}
        &=\frac{k^2 G_k}{6 \pi } \Bigg[-12 k^2-\Lambda _k+\frac{3  \left(6 k^4-27 k^2 \Lambda _k+20 \Lambda _k^2\right)}{k^2-2 \Lambda _k}\exp \left(\frac{2 \Lambda _k}{k^2-2 \Lambda _k}\right)\Bigg]\,.
        \label{eq:ptFlowLambdadimensionful}
\end{align}
}

\paragraph{Fixed points and critical exponents.}

We now write the RG equations\,\eqref{eq:ptFlowGdimensionful} and\,\eqref{eq:ptFlowLambdadimensionful} in terms of the dimensionless Newton and cosmological constants
\begin{equation}
	g_k=k^2G_k,
	\qquad
	\lambda_k={\Lambda_k\over k^2}\,,
	\label{eq:dimlessCouplings}
\end{equation}
look for the fixed points and calculate the corresponding critical exponents.

Introducing\,\eqref{eq:dimlessCouplings} in\,\eqref{eq:ptFlowGdimensionful} and\,\eqref{eq:ptFlowLambdadimensionful} we get
\begin{align}
	k\dv{g}{k}
	&=
	2g
	-
	{g^2\over 6\pi}
	\left[
	1+
	{9+12\lambda\over 1-2\lambda}
	\exp\left({2\lambda\over 1-2\lambda}\right)
	\right] \equiv \beta_g (g,\lambda)
\label{eq: dimless pt g}	\\[1ex]
k\dv{\lambda}{k}
	&=
	-2\lambda
	+
	{g\over \pi}
	\left[
	-2
	-
	{\lambda\over 6}
	+
	{6-27\lambda+20\lambda^2\over 2(1-2\lambda)}
	\exp\left({2\lambda\over 1-2\lambda}\right)
	\right]\equiv \beta_\lambda (g,\lambda)\,.
	\label{eq: dimless pt lambda}
\end{align}
The first term in each beta function is the canonical scaling contribution, while the remaining terms are the spectral quantum corrections. The non-polynomial dependence on $\lambda$ is inherited from the singular terms $k^2-2\Lambda_k$ in\,\eqref{eq:ptFlowGdimensionful},\,\eqref{eq:ptFlowLambdadimensionful} and from the smooth proper-time implementation of the shell. 

The fixed points are the solutions of $\beta_g=\beta_\lambda=0$ in the regular domain $\lambda<1/2$. Besides the Gaussian fixed point $(\lambda_*,g_*)=(0,0)$, we also find the non-trivial fixed point (first quadrant of the $(\lambda,g)$ plane)
\begin{equation}
	(\lambda_*,	g_*)\simeq(0.149,1.536)\,.
	\label{eq:NGFPPTvalues}
\end{equation}

\begin{figure}[t]
        \centering
        \includegraphics[width=0.82\textwidth]{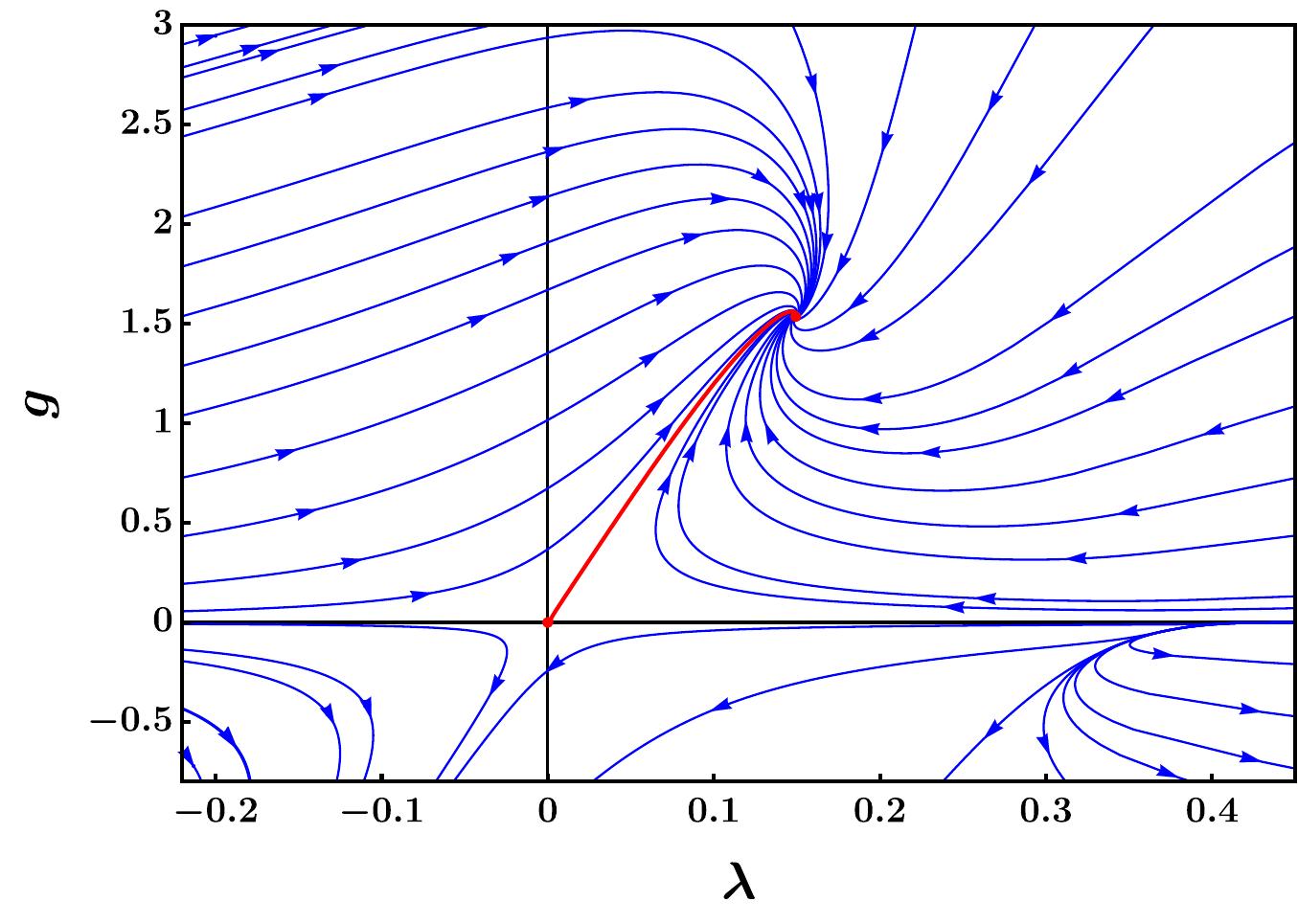}
        \caption{RG flow in the $(\lambda,g)$ plane for the smooth spectral cutoff. Red dots denote the Gaussian and the non-Gaussian fixed points. The red curve is the separatrix that connects them. The arrows point towards the UV.}
        \label{fig:flowPT}
\end{figure}

To determine the critical exponents we linearize the flow around the fixed point,
\begin{equation}
	k \dv{}{k}
	\begin{pmatrix}
		\delta\lambda\\[2pt]
		\delta g
	\end{pmatrix}
	=M
	\begin{pmatrix}
		\delta\lambda\\[2pt]
		\delta g
	\end{pmatrix},
	\qquad
	M=
	\begin{pmatrix}
		\partial_\lambda\beta_\lambda & \partial_g\beta_\lambda\\[2pt]
		\partial_\lambda\beta_g & \partial_g\beta_g
	\end{pmatrix}_{(\lambda_*,g_*)} .
	\label{eq:stabilityPT}
\end{equation}
As expected, for the Gaussian fixed point we find the canonical eigenvalues $-2$ and $+2$: the $\lambda$ axis is UV-attractive while the $g$ axis is UV-repulsive. Concerning the non-trivial fixed point\,\eqref{eq:NGFPPTvalues}, we find the eigenvalues $\alpha_{1,2}\simeq-3.194\pm 1.781\,i$ so that with the standard convention, $\theta_i=-\alpha_i$, the critical exponents are
\begin{equation}
        \theta_{1,2}\simeq3.194\mp 1.781\,i .
        \label{eq:criticalExponentsPT}
\end{equation}
We then see that the non-Gaussian fixed point\,\eqref{eq:NGFPPTvalues} is UV-attractive and the flow approaches it with a spiralling behaviour, as shown in Fig.\,\ref{fig:flowPT}.

This is one of the most interesting results of the present work. We have just seen that, implementing the Wilsonian RG flow with the help of a smooth running spectral cut, that ultimately means resorting to the eigenvalues of the covariant Laplace-Beltrami operators to introduce a physically meaningful separation of scales, the asymptotic safety scenario (at least in the Einstein-Hilbert truncation considered in the present work) is realized. In our opinion, the fact that this scenario emerges in such a framework provides support to the scenario itself. As mentioned in the Introduction, in a previous study we obtained a different result \cite{Branchina:2024lai}. The strength of the present approach stems from the identification of the running scale in terms of a spectral cut\cite{Branchina:2026vvk}.

In the next section we derive the RG equations for $G_k$ and $\Lambda_k$ defining the flow in terms of a hard spectral cutoff.

\section{Hard spectral cutoff and RG equations}
\label{sec: Hard cutoff}

Aim of the present section is to realize the same idea of a spectral cutoff in a different way. Referring as before to the general considerations developed in section\,\ref{sec:oneloop}, in particular to Eqs.\,\eqref{eq:spinScalarShift} and\,\eqref{eq:shell} and comments therein, we now implement the spectral cut replacing\,\eqref{eq:ptTrace Os} with
\begin{equation}
        \Big(\log \det \mathcal{O}_{s,\alpha}\big|_{k,k-\delta k}\Big)_{\rm hard}
        =
        \sideset{}{'}\sum_{(k -\delta k)^2 \leq \lambda_n^{(0)}\leq k^2}
        D_n^{(s)}
        \log \left(\frac{\lambda_n^{(0)}-{s/a^2}+\alpha}{k^2}\right)\,.
        \label{eq:hardTraceGeneric}
\end{equation}
The $'$ in $\sum$ indicates that the sum over the spectral shell $(k -\delta k)^2 \leq \lambda_n^{(0)}\leq k^2$ has to be taken  in such a way that discontinuities that arise at the boundaries are avoided. To this end we adopt the midpoint prescription that we describe below. 

From\,\eqref{eq:sphereSpectra}$_1$ we have that $\lambda^{(0)}_n=(n^2+3n)/a^2$.  The relation 
\begin{equation}
	\lambda^{(0)}_n={n^2+3n\over a^2}\leq k^2
	\label{eq:scalar_level_condition}
\end{equation}
indicates that the integers $n$ to be taken in the sum have to satisfy the inequality ($\lfloor x \rfloor$ is the floor function)
\begin{equation}
	n \leq \left\lfloor \frac{1}{2} \left(\sqrt{4 a^2 k^2+9}-3\right) \right\rfloor \equiv n_{\rm max}\,.
	\label{nbound}
\end{equation}
The sums of the kind\,\eqref{eq:hardTraceGeneric} that define the RG step (see below for the whole expression) are then performed up to $n_{\rm max}$. Now we rewrite $n_{\rm max}$ as\footnote{We recall that $\lfloor x \rfloor = x - \{x\}$, where $\{x\}$ is the fractional part of $x$.}
\begin{equation}
	n_{\rm max} = \frac{1}{2} \left(\sqrt{4 a^2 k^2+9}-3\right) - \left\{\frac{1}{2} \left(\sqrt{4 a^2 k^2+9}-3\right)\right\}\,.
\end{equation}
The midpoint prescription consists in replacing the fractional part with its average $\langle \{x\} \rangle = 1/2$, i.e.
\begin{equation}
	n_{\rm max} = \frac{1}{2} \left(\sqrt{4 a^2 k^2+9}-3\right) - \left\{\frac{1}{2} \left(\sqrt{4 a^2 k^2+9}-3\right)\right\} \rightarrow  \frac{1}{2} \left(\sqrt{4 a^2 k^2+9}-3\right) - \frac 12\,.
\end{equation}
Similar considerations hold for the other inequality, namely $(k -\delta k)^2 \leq \lambda_n^{(0)}$. This procedure amounts to smoothing the edge ambiguities while preserving the hard character of the spectral cutoff. 

Replacing now each of the determinants in\,\eqref{eq:oneloopdet} according to\,\eqref{eq:hardTraceGeneric}, expanding to first order in $\delta k$, and taking the limit $\delta k\to 0$, we obtain for the running action $S_k$ the flow equation (as before in the right-hand side the derivatives with respect to $k$ are taken keeping $G_k$ and $\Lambda_k$ fixed)
\begin{align}
	k{\partial S_k\over\partial k}
	=-\frac12 k
	\,{\partial\over\partial k}
	\Bigg[\,\,
	& \sideset{}{'}\sum_{ \lambda_n^{(0)}\leq k^2}
	D_n^{(0)}
	\log\left(\frac{\lambda_n^{(0)}-2\Lambda_k}{k^2}\right)+ \sideset{}{'}\sum_{ \lambda_n^{(0)}\leq k^2}
	D_n^{(0)}
	\log\left(\frac{\lambda_n^{(2)}+6/a^2-2\Lambda_k}{k^2}\right) \nonumber \\
	- & \sideset{}{'}\sum_{ \lambda_n^{(0)}\leq k^2}
	D_n^{(1)}
	\log\left(\frac{\lambda_n^{(0)}-4/a^2}{k^2}\right) - \sideset{}{'}\sum_{ \lambda_n^{(0)}\leq k^2}
	D_n^{(0)}
	\log\left(\frac{\lambda_n^{(0)}-6/a^2}{k^2}\right)\Bigg]
	\label{eq:hcRG}
\end{align}
Inserting in the left-hand side of\,\eqref{eq:hcRG} the running action $S_k^{(a)}$ given in\,\eqref{eq:EHtruncationRunningPT}, expanding the right-hand side of\,\eqref{eq:hcRG} for large $a$, and matching the coefficients of $a^2$ and $a^4$, we get the flow equations for $1/G_k$ and $\Lambda_k/G_k$
\begin{align}
        k\dv{}{k}
        \left({1\over G_k}\right)
        &=
        {203k^6-641k^4\Lambda_k+482k^2\Lambda_k^2
        \over
        36\pi\left(k^2-2\Lambda_k\right)^2}
        -{117k^2\over72\pi}
        \log\left(1-{2\Lambda_k\over k^2}\right)
        \label{eq:hardFlowInvG} \\
        k\dv{}{k}
        \left({\Lambda_k\over G_k}\right)
        &={k^4\over2\pi}
        \left[
        1-6\log\left(1-{2\Lambda_k\over k^2}\right)
        \right] .
        \label{eq:hardFlowLambdaOverG}
\end{align}
As for the RG equations\,\eqref{eq:ptFlow1overG} and\,\eqref{eq:ptFlowLambdaOverG} of the previous section (smooth spectral cutoff), Eqs.\,\eqref{eq:hardFlowInvG} and\,\eqref{eq:hardFlowLambdaOverG} display a singular behaviour for $k^2 \to 2\Lambda_k$. They are then valid only for $\Lambda_k < k^2/2$. Concerning the presence in \eqref{eq:hardFlowInvG} and \eqref{eq:hardFlowLambdaOverG}  of terms that depend on $1 - \frac{2\Lambda_k}{k^2}$, the same considerations made below Eqs.\,\eqref{eq:ptFlow1overG} and \eqref{eq:ptFlowLambdaOverG} hold. As said in the previous section,  their presence is  related to the well-known IR instability of the propagator in de Sitter space. 
 
Writing the equations for $G_k$ and $\Lambda_k$ we obtain
{\small
\begin{align}
        &k{\dd G_k\over\dd k}
        =-\frac{G_k^2 k^2}{36 \pi} \left( {203k^4-641k^2\Lambda_k+482\Lambda_k^2
        	\over
        	\left(k^2-2\Lambda_k\right)^2}
        -{117\over2}
        \log\left(1-{2\Lambda_k\over k^2}\right)\right),
        \label{eq:hardFlowGdimensionful}\\
        &k{\dd \Lambda_k\over\dd k}
        =\frac{k^4 G_k}{2 \pi } \left[ 1-6 \log \left(1-\frac{2 \Lambda _k}{k^2}\right)+\frac{\Lambda _k}{36 k^2}  \left(\frac{3 k^2 \left(106 \Lambda _k-55 k^2\right)}{\left(k^2-2 \Lambda _k\right){}^2}-241+117 \log \left(1-\frac{2 \Lambda _k}{k^2}\right)\right)\right]\,.
        \label{eq:hardFlowLambdadimensionful}
\end{align}
}

\paragraph{Fixed points and critical exponents.}

We move on to write the dimensionless RG equations for $g_k = k^2 G_k$ and $\lambda_k = \Lambda_k/k^2$ and study the fixed points of the system.

Writing equations \eqref{eq:hardFlowGdimensionful} and \eqref{eq:hardFlowLambdadimensionful} in terms of their dimensionless counterpart, we have 
\begin{align}
k\dv{g}{k}&=2g-\frac{g^2}{36 \pi}\left[{203-641\lambda+482\lambda^2
\over \left(1-2\lambda\right)^2}-{117\over2}\log\left(1-2\lambda\right)\right]\equiv \beta_g(g,\lambda),\\[1ex]
k\dv{\lambda}{k}&=-2\lambda+{g\over 2\pi}\left[{18-275\lambda+713\lambda^2-482\lambda^3\over18\left(1-2\lambda\right)^2}+\left({13\lambda\over 4}-6\right)\log\left(1-2\lambda\right)\right] \equiv \beta_\lambda(g,\lambda).
\end{align}
Comments similar to those below Eqs.~\eqref{eq: dimless pt g} and \eqref{eq: dimless pt lambda} apply here. Again we observe that these equations are valid only in the domain where $\lambda<1/2$. 

Besides the Gaussian fixed point 
\begin{equation}
(\lambda_*,g_*)=(0,0),
\end{equation}
these equations possess two non-trivial fixed points. They are 
\begin{equation}
(\lambda^{(1)}_{*},g^{(1)}_{*})\simeq (0.080,0.985)
\label{eq:HardASvalues}
\end{equation}
and
\begin{equation}
(\lambda^{(2)}_*,g^{(2)}_*)\simeq (-2.168,6.022).
\label{eq:HardNegativeValues}
\end{equation}

We determine the critical exponents by linearizing the flow around the fixed points. As before, for the Gaussian fixed point the $\lambda$ axis is UV-attractive while the $g$ one is UV-repulsive. For $(\lambda^{(1)}_{*},g^{(1)}_{*})$, solving for the eigenvalues of the stability matrix we find that the critical exponents are 
\begin{equation}
	\theta^{(1)}_{1,2}
	\simeq2.015\mp0.734\,i \,,
	\label{eq:criticalExponentsHardAS}
\end{equation}
while for $(\lambda^{(2)}_{*},g^{(2)}_{*})$ we find
\begin{equation}
	\theta_1^{(2)}\simeq-2.697,
	\qquad
	\theta_2^{(2)}\simeq3.356 .
	\label{eq:criticalExponentsHardNegative}
\end{equation}

\begin{figure}[t]
	\centering
	\includegraphics[width=0.82\textwidth]{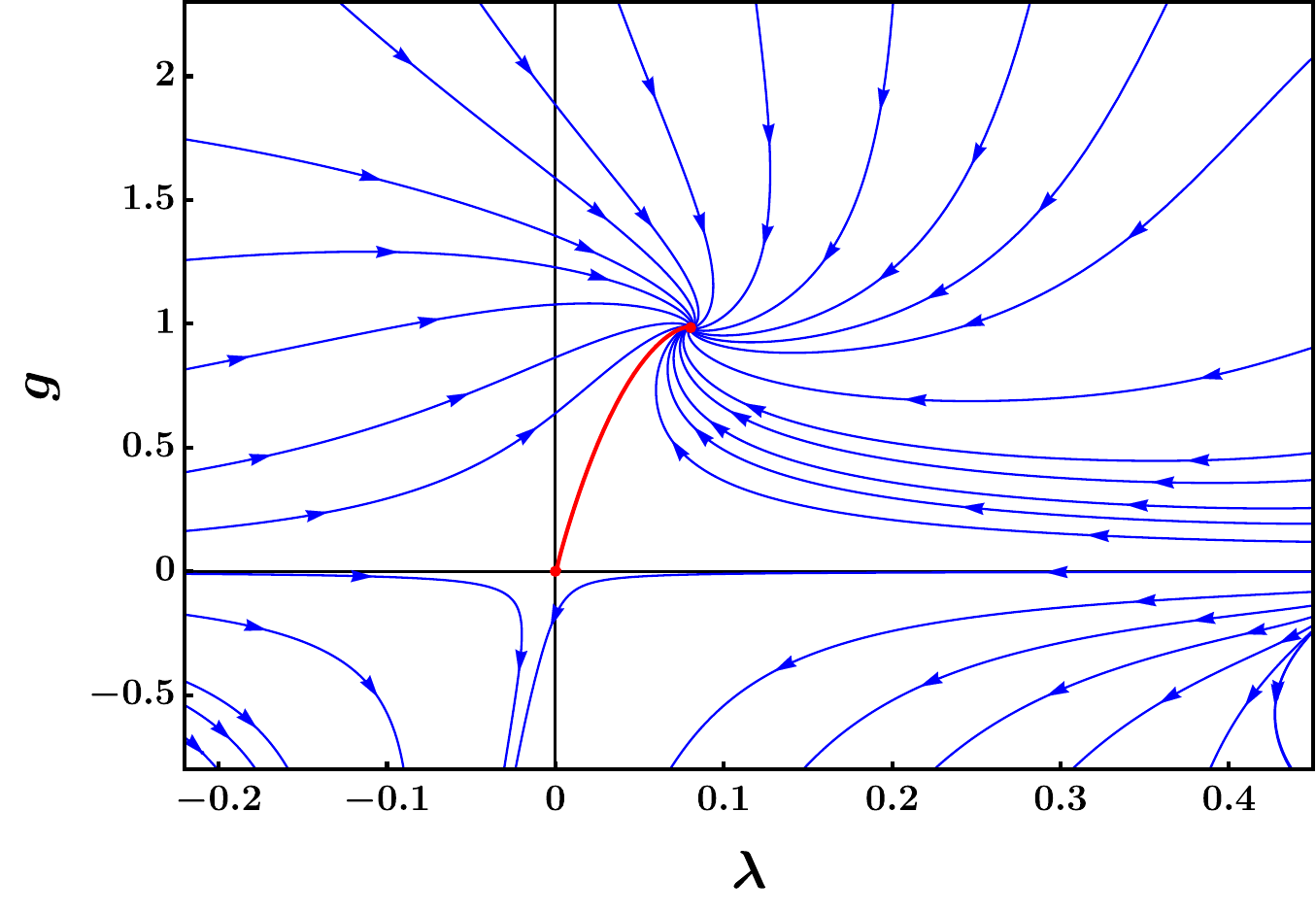}
	\caption{RG flow for the case of the hard spectral cutoff in a region of the $(\lambda,g)$ plane containing only the Gaussian fixed point and the UV-attractive one. The graphical conventions are as in Fig.~\ref{fig:flowPT}.}
	\label{fig:flowHardAS}
\end{figure}

\begin{figure}[t]
        \centering
        \includegraphics[width=0.82\textwidth]{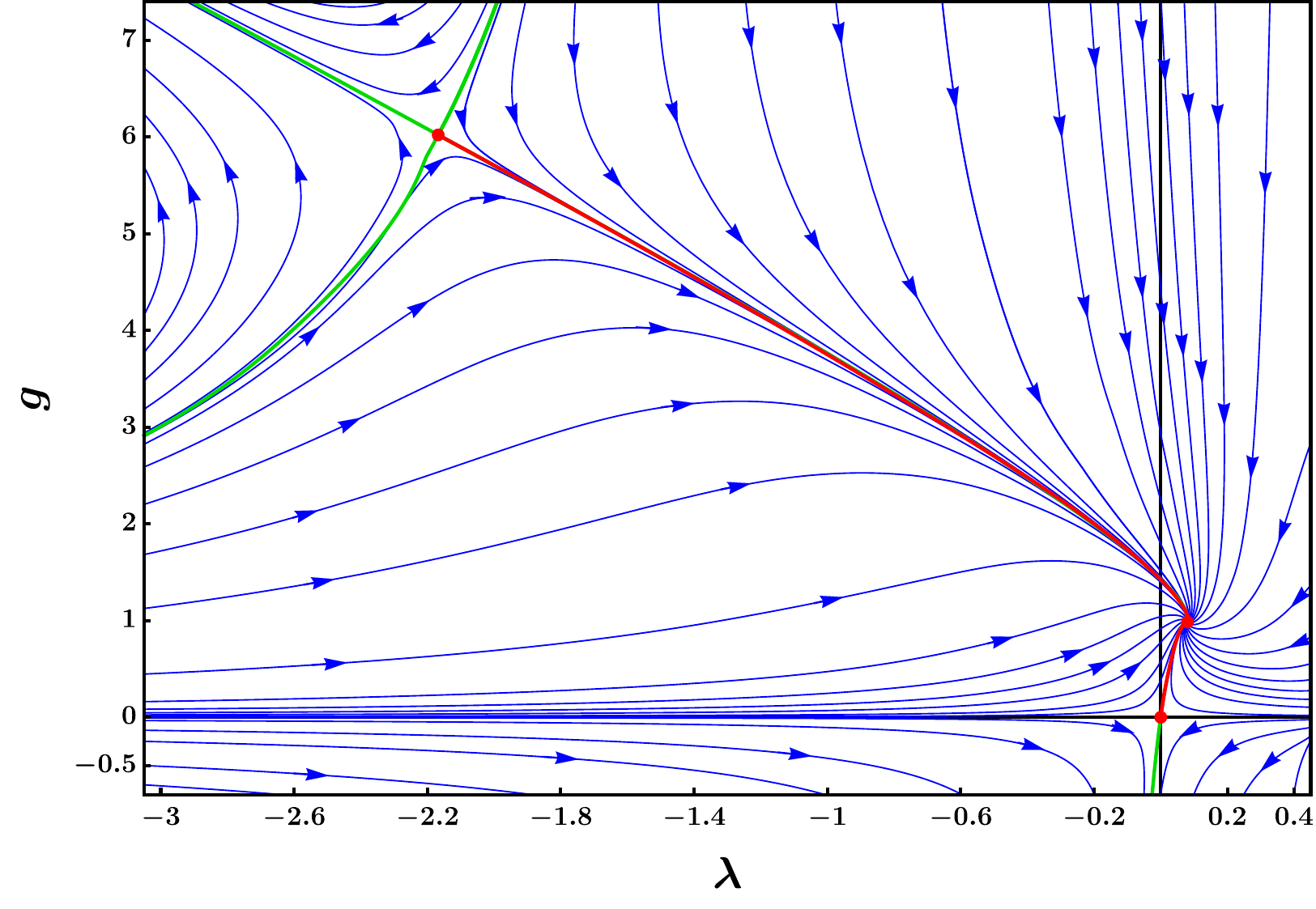}
        \caption{RG flow in the $(\lambda,g)$ plane for the hard spectral cutoff in a region larger than the one considered in Fig.~\ref{fig:flowHardAS}. The presence of the third fixed point $(\lambda^{(2)}_{*},g^{(2)}_{*})$ and its impact on the flow are shown.}
        \label{fig:flowHardNegative}
\end{figure}

These results show that, as for the smooth case, the hard implementation of the spectral running cutoff displays the presence of a non-Gaussian UV-attractive fixed point. Again, the asymptotic safety scenario with a UV-flow that approaches $(\lambda^{(1)}_{*},g^{(1)}_{*})$ by circling around it is realized. This is shown in Fig.\,\ref{fig:flowHardAS}.
Concerning the fixed point $(\lambda^{(2)}_{*},g^{(2)}_{*})$, Eq.\,\eqref{eq:criticalExponentsHardNegative} shows that it has one UV-attractive and one UV-repulsive direction.  It is then a saddle (not a UV-attractor) that influences the flow in the region of the plane with negative values of $\lambda$. This behaviour is displayed in Fig.\,\ref{fig:flowHardNegative}.

Asymptotic safety is thus found even when the spectral cutoff is implemented in terms of a hard cut. We think it worth to repeat here that the fact that such a behaviour arises from a spectral implementation of the running scale \cite{Branchina:2026vvk}, i.e.\,from a separation of scales obtained resorting to the covariant Laplace-Beltrami operator, seems to provide further support to the AS scenario. 

\section{Conclusions and outlooks}
\label{sec:conclusions}

According to the analysis that we performed in \cite{Branchina:2026vvk}, in this work we have implemented the Wilsonian RG paradigm in quantum gravity introducing a covariant spectral running cutoff, considering the Einstein-Hilbert truncation. The construction is based on the observation that the scales, and a fortiori the RG step, are set by the spectrum of the covariant Laplace-Beltrami operator. The Wilsonian separation between lower and higher energy scales emerges as the direct and natural consequence of the invariant separation of the eigenvalues of the Laplacian. 

We considered two realizations of this guiding principle. In the first one we used a smooth spectral cutoff. For the second one we considered a sharp implementation. In both cases we found that the asymptotic safety scenario is realized. More precisely, with the 
smooth spectral flow we found a Gaussian fixed point and a non-Gaussian UV-attractive fixed point (with a couple of complex conjugate critical exponents). For the case of the hard spectral cut, in addition to the Gaussian and the UV-attractive non-Gaussian fixed point, we found a third fixed point with one UV-attractive and one UV-repulsive eigendirection (with $\lambda_* <0$). The presence of this additional fixed point, however, does not change the UV behaviour of the theory in the $\lambda>0, \, g>0$ region, where the asymptotic safety scenario is realized. In the past we got a different result, according to which it seemed that the RG flow did not present the  AS scenario \cite{Branchina:2024lai}. As compared to this previous work, the new crucial element that allowed to get the results of the present work is the identification of the running scale in terms of a spectral cut, that follows from the analysis that we presented in  \cite{Branchina:2026vvk}. 

Our present construction differs from the usual realizations of the AS scenario. In the case of the implementation of the RG flow in the framework of the effective average action formalism, for instance, the latter is determined  by the trace of $(\Gamma_k^{(2)}+R_k)^{-1}k\partial_k R_k$, where $R_k$ is a regulator function ($\Gamma_k$ is the average action at the scale $k$ and $\Gamma_k^{(2)}$ the corresponding inverse propagator). 
Our implementation  refers directly to the invariant eigenvalues (scales) of the Laplace-Beltrami operator: the running scale $k$ itself is determined according to the inbuilt separation of said eigenvalues. The difference  does not merely amount to a technical change of regulator. The novelty of the present work is the application of the idea first introduced in\,\cite{Branchina:2026vvk} (spectral cut) to the RG analysis of quantum gravity. 

It is worth to come back here to the point central to the analysis developed in the present work, a point that we have stressed  in\cite{Branchina:2026vvk} and all along this work. The Wilsonian paradigm in  gravity (in gauge theories) is naturally implemented  through the introduction of a \vv spectral cutoff'' $k$ on the eigenvalues of the covariant Laplacians.  In these cases,  the  separation of scales cannot be obtained with a momentum cutoff since the latter does not respect gauge/diffeomorphism invariance. 

The analysis of the present work represents a first step and we believe that several developments are possible. Naturally, it would be interesting to use this approach to go beyond the Einstein-Hilbert truncation including higher-curvature operators, such as  $R^2$, $R_{\mu\nu}R^{\mu\nu}$, ... .  Another direction could be to extend the analysis considering the inclusion of matter fields. Also, one could try to repeat the analysis on backgrounds different from the four-sphere, in order to disentangle projection effects from genuine spectral-cutoff effects.

We hope that the methods introduced in the present work might be of help in studying (at least some of) the open questions in quantum gravity (in particular in the AS scenario), and more generally in implementing the Wilsonian paradigm in those cases where the separation of scales is naturally realized through the eigenvalues of the Laplacian operators that enter the quantum fluctuation determinants.  

\section*{Acknowledgments}

The work of VB, FC, RG and AP is carried out within the INFN project  QGSKY.

\end{document}